\documentclass[aps,preprint,prb,showkeys,showpacs]{revtex4}
\usepackage{epsfig}
\usepackage{hyperref}

\begin{document}

\def\il{I_{low}}
\def\iu{I_{up}}
\def\eeq{\end{equation}}
\def\ie{i.e.}
\def\etal{{\it et al. }}
\def\prb{Phys. Rev. { B}}
\def\pra{Phys. Rev. { A}}
\def\prl{Phys. Rev. Lett. }
\def\pla{Phys. Lett.}
\def\pb{Physica B}
\def\ajp{Am. J. Phys. }
\def\epjb{Eur. Phys. J B }

\def\apl{Appl. Phys. Lett. }
\def\jpc{J. Phys. C }
\def\rmp{Rev. of Mod. Phys. }
\def\jap{J. Appl. Phys. }
\def\mpl{Mod. Phys. Lett. { B}}
\def\ijmp{Int. J. Mod. Phys. { B}}
\def\ijp{Ind. J. Phys. }
\def\ijpap{Ind. J. Pure Appl. Phys. }
\def\ibmjrd{IBM J. Res. Dev. }
\def\pjp{Pramana J. Phys.}
\def\pm{Philos. Mag. }
\def\ssc{Solid State Commun.}

\title{A comparative study of some models of incoherence at the
  mesoscopic scale.}
\author{Colin Benjamin}
\email{colin@iopb.res.in}
\affiliation{Institute of Physics, Sachivalaya Marg, Bhubaneswar 751 005,
  Orissa, India}
\author{A. M. Jayannavar}
\email{jayan@iopb.res.in}
\affiliation{Institute of Physics, Sachivalaya Marg, Bhubaneswar 751 005,
  Orissa, India}

\date{\today}

\begin{abstract}
  
  The pre-existing literature on phenomena at the mesoscopic scale is
  concerned among other things with phase coherent transport. Phase
  coherent transport dominates at very low temperatures. With increase
  in temperature, as the system size becomes comparable to the
  inelastic mean free path phase incoherence sets in. This incoherence
  further leads to dephasing, and as a consequence purely quantum
  effects in electron transport give way to classical macroscopic
  behavior. In this work we consider two distinct phenomenological
  models of incoherent transport, the Coherent Absorption and Wave
  Attenuation models. We reveal some physical problems in the Coherent
  Absorption model as opposed to the Wave Attenuation model.  We also
  compare our proposed model with experiments in case of the much
  studied peak to valley ratios in resonant tunneling diodes,
  magneto-conductance oscillations and Fano resonances in case of
  Aharonov-Bohm rings.

\end{abstract}



\maketitle 

\section{Introduction}

Absorption of particles has been long studied in quantum
mechanics\cite{gas}. In many quantum mechanical scattering experiments
there is absorption of the incident beam; the target particle may get
excited, or change it's state or another particle may emerge (if it
isn't a fundamental particle). Here Absorption implies inelastic
scattering which can be simulated phenomenologically in the
Schroedinger equation by means of a complex potential, sometimes
called optical potential in nuclear physics
literature\cite{schiff,ferry,stone}. The Hamiltonian becomes
non-hermitian and there is no particle number conservation. Hence
forth known as Coherent Absorption.  The method of Coherent Absorption
has also been utilized to study the localization of waves in a random
media in presence of absorbing/amplifying medium\cite{pradhan}. It
should be noted that absorption should not be confused with
disappearance of particles, which is unacceptable.  However absorption
should be treated as particles that have entered into different
inelastic channels, or loss of particles from the coherent channel.
This is not the only way in which absorption induced inelastic
scattering can be simulated. The other method is Wave Attenuation,
which has been earlier utilized to study problems relating to 1D
localization\cite{joshi} and dephasing of Aharonov-Bohm
oscillations\cite{colin}. Both of these models have been used to
calculate quantum mechanically the conditional sojourn times taken by
a particle to traverse a region of interest\cite{soj1,soj2}. In the
not so distant past, some experimental studies\cite{neutron} were
performed on neutron interferometry to bring out the effect of
different types of absorption. In section II, we give brief
definitions and also bring out the differences in the above
phenomenological models of incoherence.

The reason we are interested in absorption in mesoscopic systems is
due to the fact that incoherent transport plays an important role in
these systems. Of course, Landauer has demonstrated\cite{land} since
long that the conductance of these systems, with elastic scattering
alone, can be obtained from the transmission probability of the
sample, that is it's overall scattering matrix. In a realistic model,
however, phase breaking processes have to be taken into account which
destroy the quantum coherence and lead to dissipation\cite{imry}. Such
effects can be phenomenologically included into the Landauer picture
by the models of absorption considered above.

\section{Models of Incoherence}

Incoherence in particle transport can be modeled in several ways.
Among the first was by B\"{u}ttiker \cite{buti,buti1} who considered
an electron reservoir coupled by a lead to a mesoscopic system as a
phase breaker or inelastic scatterer (voltage probe). This approach
has been widely used to investigate the effect of incoherence or
dephasing on the conductance. This method which uses voltage probes
as dephasor's is interesting because of it's conceptual clarity and
it's close relation to experiments. It provides a useful trick to
simulate lack of full coherence in transport properties. This method
of addressing the problem of dephasing has the advantage that
inelastic phase randomizing processes can be incorporated by solving
an elastic time independent scattering problem. Beyond B\"{u}ttiker's
model, Coherent Absorption \cite{ferry,jayan} and Wave Attenuation
models \cite{colin,joshi} have also been used to simulate dephasing.
However, in the aforesaid models energy relaxation and thermal
effects\cite{mortensen} are ignored. Thermal effects can be
incorporated by taking into account thermal distribution (Fermi-Dirac
function) of electrons. In mesoscopic systems, transmission functions
are more often than not constant over the energy range wherein
transport occurs (at low temperatures) and one can ignore energy
relaxation or ``vertical flow''\cite{datta} of electron carrier's in
these systems. Brouwer and Beenakker have corrected some of the
problems associated with voltage probe and Coherent Absorption models,
(see Refs.[\onlinecite{brouwer,colin}] for details), and given a
general formalism for calculating the conductance(G) in the presence
of inelastic scattering. Furthermore, methods based on Coherent
Absorption and Wave Attenuation can make use of this above formalism.

In this work we compare the method of Coherent Absorption and Wave
Attenuation models. For this we study some topics in mesoscopic
physics wherein to model inelastic scattering recourse to such models
of absorption is often implied. Two representative topics wherein we
apply these absorptive models are Resonant tunneling diodes(RTD) and
Aharonov-Bohm(AB) rings.  In RTD's, the discrepancy between
experimental and theoretical peak to valley ratios has often been
attributed to incoherent or sequential transport\cite{gaston,rubio}.
We model incoherent transport through both of these models of
absorption and point out certain problems with the Coherent Absorption
model as compared to the Wave Attenuation model. In AB rings also the
same anomalies with regard to the Coherent Absorption model are put
forth through some representative plots of conductance.  Some aspects
of dephasing on AB oscillations are discussed. Further we utilize the
method of Wave Attenuation to model incoherent transport and explain
the transition of Fano to Lorentzian line shapes obtained in the
Conductance of an Aharonov-Bohm ring with a symmetrically placed
quantum dot on it's upper arm.

\section{Resonant tunneling diodes}

\begin{figure}[h]
\protect\centerline{\epsfxsize=3.0in \epsfbox{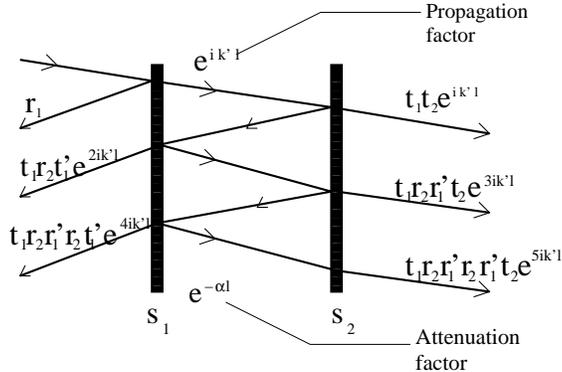}}
\caption{Summing the different paths, $S_1$ and $S_2$ denote the two
  scatterer's. $l$ is the distance between them. $e^{i k^{\prime}l}$
  and $e^{-\alpha l}$ denote the propagation and attenuation factors
  in the locality of interest.}
\end{figure}

Technological progress in semiconductor fabrication has led to the
possibility of devices which are based on quantum effects. A simple
such device is the double barrier diode\cite{esaki,chang,sollner},
analogous to the model of a Fabry-Perot interferometer as in FIG.~1.
It consists of two potential barriers separated by a thin quantum
well.  The transmission of electrons across this diode shows sharp
resonances when energy of incident electrons are close to that of a
quasi bound state. FIG.~2 shows a double barrier diode wherein a
quantum well (region III) is sandwiched between two equal barrier's II
and IV.  Region's I and V denote electrodes. Here incident electrons
are described by plane waves. They tunnel through the left and right
barrier's via the quantum well. The potential felt by the electron in
various regions is depicted in FIG.~2. A finite voltage bias $V$ is
applied across the system. The electron wave in the well experiences
multiple reflections due to the presence of barriers (see FIG.~1) and
then tunnels out through the right barrier. To calculate the
transmission and reflection coefficients we use the transfer matrix
method\cite{Yu}.  The transfer matrices of either of the barrier's are
calculated by matching the wave functions and their derivatives at the
boundaries. To calculate the transfer matrices we choose the II/III
interface as the origin of the coordinate system, and the transfer
matrix of the first barrier is given by-

\begin{figure}[h]
  \protect\centerline{\epsfxsize=3.0in \epsfbox{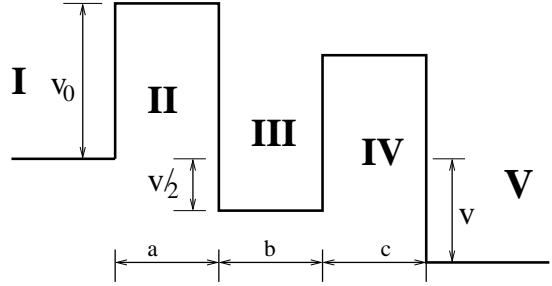}}
\caption{The double barrier heterostructure.}
\end{figure}

\[M=\left(\begin{array}{cc}
w& z^{*}\\
z& w^{*}\\
\end{array} \right) \]

with $w=\frac{e^{i k_{1} a}}{2}((1+f_{13})cosh(K_{2}
a)-i(f_{23}-f_{12})sinh(K_{2} a))$, and $z=\frac{e^{-i k_{1}
    a}}{2}((1-f_{13})cosh(K_{2} a)-i(f_{23}+f_{12})sinh(K_{2} a))$,
wherein $f_{13}=k_{3}/k_{1}$, $f_{23}=k_{3}/K_{2}$,
$f_{12}=K_{2}/k_{1}$, $K_{2}=\sqrt{2m^{*}(V_{0}- E)/\hbar^{2}}$, and
$k_{1}=\sqrt{2m^{*}E/\hbar^{2}}$.  We model incoherent transport only
in the well. In the method of Wave Attenuation 
$k_{3}=\sqrt{2m^{*}(E+V/2)/\hbar^{2}}$ and we use wave attenuation
factor to simulate dephasing as explained below. In the case of
Coherent Absorption, the potential in the well is complex and $k_{3}=k_{p}+i
k_{m}=\sqrt{2m^{*}(E+V/2-iV_{i})/\hbar^{2}}$.  Here $V_i$ is the
imaginary part of the potential which is characterized as an incoherent
parameter. After obtaining the transfer matrix the S-Matrix
\cite{griffith}of the barrier is calculated by using the formula-

\[S=(\frac{1}{M_{11}})\left(\begin{array}{cc}
M_{21}& det M\\
1& -M_{12}\\
\end{array} \right) \]
wherein M is the transfer matrix of the barrier and indices denote row
and column.

The S-Matrix of the second barrier is analogously written.  The
S-Matrix of the well is -
\[S_{w}=\left(\begin{array}{cc}
e^{i k_{3} b}e^{-\alpha b}&0\\
0& e^{i k_{3} b}e^{-\alpha b}\\
\end{array} \right) \]

in presence of Wave Attenuation, while in presence of coherent
absorption \cite{jayan}the S-Matrix of the well is -
\[S_{w}=\left(\begin{array}{cc}
e^{i k_{3} b}&0\\
0& e^{i k_{3} b}\\
\end{array} \right) \]
with $k_{3}=k_{p}+i k_{m}=\sqrt{2m^{*}(E+V/2-i V_{i})/\hbar^{2}}$, and
$k_{p}, k_{m}$ are defined as above.  In the Wave Attenuation method,
an attenuation constant per unit length is introduced in the well,
i.e., the factor $e^{-\alpha b}$ in the free propagator amplitudes,
every time the electron\cite{colin,datta} traverses the well of the
RTD (see Fig.~1).  $b$ is the length of the quantum well. This
attenuation factor does not affect the scattering properties of the
two neighboring scatterer's as opposed to the Coherent Absorption
model. The attenuation constant $\alpha$ represents the incoherence
parameter.

Physically the model structure depicted in FIG.~2, shows sharp
resonances when the energy of the incident particles is close to that
of the quasi-bound state\cite{azbel}. In practice the barrier's are
usually of $Al_{x}Ga_{1-x}As$ and well is of $Ga As$. The important
thing realized in such structures is the concept of negative
differential resistance(NDR). This is of significance in making oscillator
components. But, theoretical models which have tried to model
experimental curves have had limited success. Two problems which have
hitherto still remained unexplained are - (1). The experimental peak
to valley ratios are far below the expected values even in good
materials, and (2). time required to store enough carrier's in the
well to reach resonant regime seems too long to be compatible with
high frequency NDR actually observed\cite{gaston}.

To explain these differences recourse has been made to the effects of
inelastic scattering\cite{stone,grincwajg}. Inelastic scattering is
inherent in these structures such as due to phonons, interfaces and
other impurities.  Modeling inelastic scattering by the models
considered above is a convenient way out from the many body treatment
needed to explain fully the differences. In case of Wave Attenuation
\cite{colin} the absorption coefficient $\alpha$ acts as a parameter
of incoherent scattering while in case of Coherent
Absorption\cite{zohta} the absorptive potential $V_i$ plays such a
role. However, it should be noted that unlike in optics where the
number of photons is not conserved so that photons can be truly
absorbed and converted into heat. In case of electrons the number of
particles has to be conserved, so $\alpha/V_{i}\neq 0$ does not mean that
electrons are absorbed but that they are scattered into different
energy channels. Thus the way these are re-injected back into the
system so as to preserve current conservation becomes important. The
coherent transmission in presence of both Coherent Absorption and Wave
Attenuation can be calculated with ease using the S-Matrices depicted
above. To calculate the incoherent contribution there are two methods.
In the method due to Zohta and Ezawa\cite{zohta} the total
transmission is defined after re-injection as the sum of two
contributions, one due to the coherent part and the other due to the
incoherent part, i.e., $T_{tot}=T_{coh}+T_{incoh}$. The incoherent
part is calculated as $T_{incoh}=\frac{T_{r}}{T_{l}+T_{r}}A$, herein
$T_{r}$ and $T_{l}$ are the probabilities for right and left
transmission from the region of inelastic scattering and $A$ is the
absorbed part which is given by $A= 1-T_{coh}-R_{coh}$. The quantities
$T_{coh}$ and $R_{coh}$ are the coherent transmission and reflection
probabilities, respectively.  This model has been used by several
other authors as well to simulate inelastic
scattering\cite{Yu,ferry,stone}.  However, modeling electron transport
by this method did not lead to satisfactory answers as the Onsager's
symmetry is shown to be violated\cite{pareek} for the two probe
conductance. In the second method due to Brouwer and
Beenakker\cite{brouwer}, the incoherent transmission is calculated
from B\"{u}ttiker's voltage probe model\cite{buti}, by mapping the three
probe B\"{u}ttiker's method into a two terminal geometry, and then
eliminating the transmission coefficients which explicitly depend on
the third probe with the help of unitarity of the S-matrix. They
considered a three terminal geometry in which one of the probes is
used as a voltage probe.  A current $I=I_1=-I_2$ flows from source to
drain. In this model, a fictitious third lead connects the system to a
reservoir at chemical potential $\mu_3$ in such a way that no current
is drawn $(I_3=0)$.  The $3X3$ unitary S-matrix of the entire system can be
written as-

\[S=\left(\begin{array}{ccc}
r_{11}& t_{12}&t_{13}\\
t_{21}& r_{22}&t_{23}\\
t_{31}& t_{32}&r_{33}
\end{array} \right) \]
Application of the relations\cite{buti,brouwer,datta}-
$I_p=\sum_q G_{pq}[\mu_p-\mu_q], p=1,2,3$ and
$ G_{pq}=(2e^2/h)T_{pq}$. 
yields the (dimensionless) two probe conductance
$G=\frac{h}{2e^2}\frac{I}{\mu_1-\mu_2}$,

\begin{eqnarray}
G=T_{21}+\frac{T_{23}T_{31}}{T_{31}+T_{32}}
\end{eqnarray}
 The transmission coefficients $T_{pq}$ are
related to the elements of the S-Matrix, $T_{pq}=|t_{pq}|^{2}$. 

On elimination of the transmission coefficients which involve the
voltage probe using the unitarity of the S - Matrix leads to\cite{brouwer}

\begin{eqnarray}
G=T_{21}+\frac{(1-R_{11}-T_{21})(1-R_{22}-T_{21})}{1-R_{11}-T_{21}+1-R_{22}-T_{12}}.
\end{eqnarray}

\begin{figure*}
\protect\centerline{\epsfxsize=6.5in \epsfbox{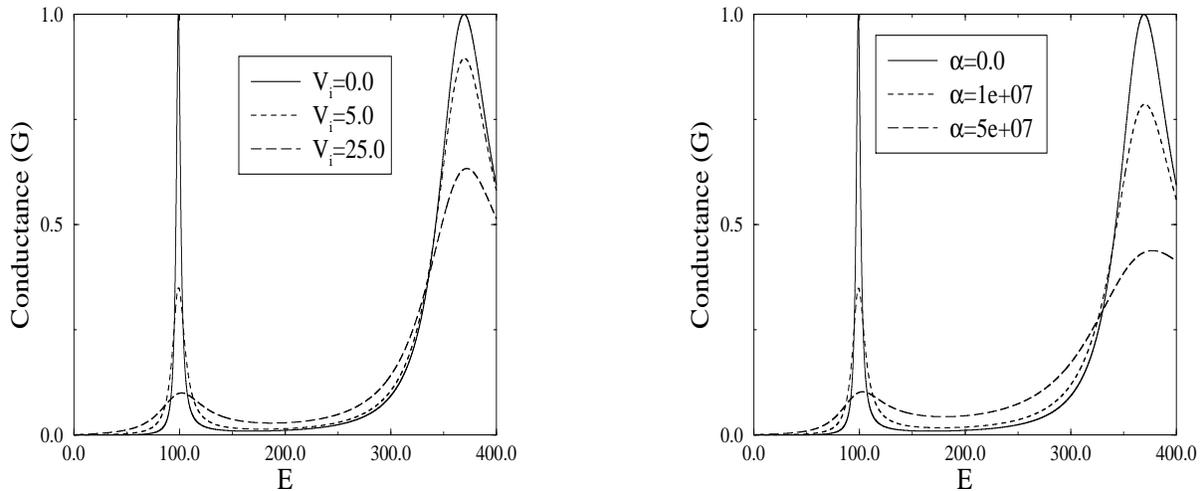}}
\caption{The total transmission (G) Vs. incident electron energy $E$
  (in mev) of a constant imaginary potential(in $mev$)/attenuation
  factor(in $m^{-1}$) in the well, for barrier heights of 0.4 ev and
  barrier and well widths of 25\AA and 50\AA.}
\end{figure*}

Here, $G$ in dimensionless form just denotes the total transmission
$T_{tot}$ in accordance with Landauer's formalism, which views
Conductance as Transmission\cite {lan_romp}.  $T_{12}$ and
$R_{22}$ are coherent transmission and reflection when carrier's are
incident from the right.  Now all the above coefficients are built
from the 2X2 S-Matrix. 
\[S^{\prime}=\left(\begin{array}{cc}
r_{11}& t_{12}\\
t_{21}& r_{22}\\
\end{array} \right) \]
The above matrix $S^{\prime}$ is a sub matrix of the 3X3 unitary
S-Matrix of the three port system. Hence matrix $S^{\prime}$ needn't
be unitary, i.e., it can represent the matrix of the non-hermitian or
absorbing system.  The first term in Eq.~2 represents the conductance
contribution from the phase coherent part. The second term accounts
for electrons that are re-injected from the phase breaking reservoir,
thereby ensuring particle conservation. This part also can be
identified as a contribution to the conductance arising from the
incoherent processes.  Eq.~2 has been obtained by a simple model for
reinjecting carrier's which conserves current only in a global way.
Another interesting approach may be to reinject carrier's based on the
local density of states\cite{butiparama}. For this one has to attach a
voltage probe at every point of the system wherein we want to depict
incoherent processes. This local approach will be more realistic and
will be dealt with in future.

\begin{figure*}
\protect\centerline{\epsfxsize=6.0in \epsfbox{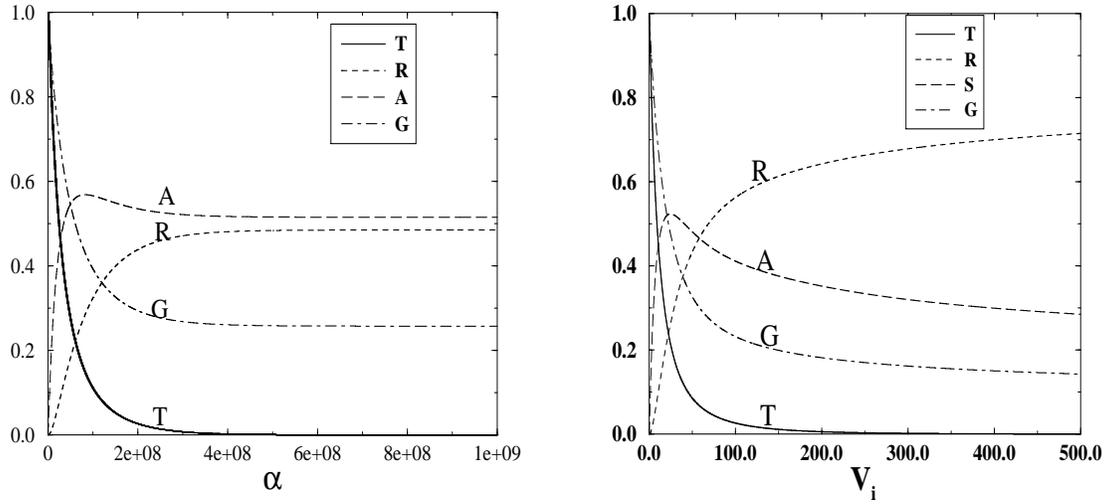}}
\caption{The Transmission(T), Reflection(R), Absorption(A) and total
  transmission (G) coefficients at resonant energy ($E_{r}$) Vs.
  strength $V_i$ (in mev) or  $\alpha$ (in $m^{-1}$) in the well, for
  barrier heights of 0.4 ev and barrier and well widths of 10\AA and
  50\AA. For this system $E_r$=88.76 mev and applied bias $V=0.0ev$.}
\end{figure*}

\begin{figure*}
\protect\centerline{\epsfxsize=6.0in \epsfbox{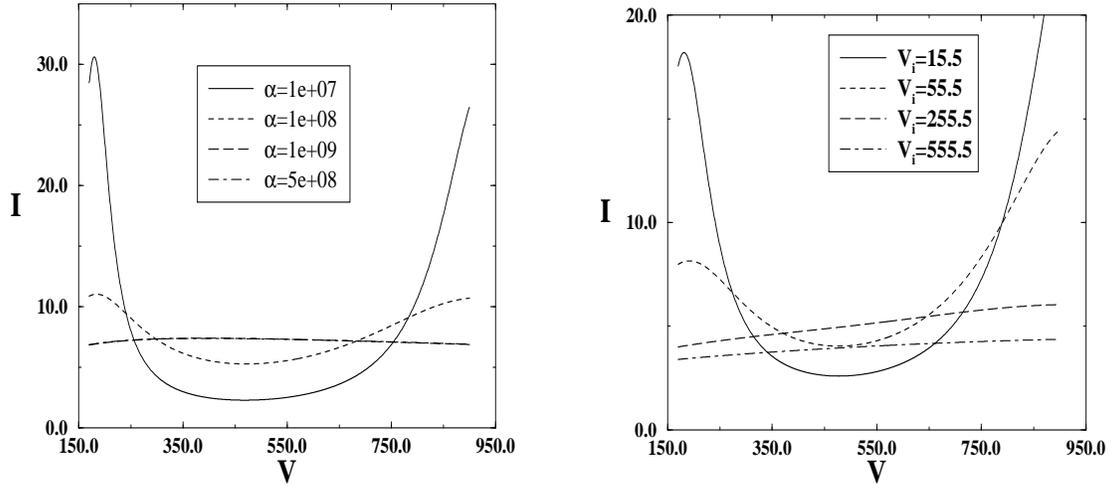}}
\caption{Tunneling current(I)  Vs. applied
  bias(V in $mev$) with increase in incoherence imaginary potential
  ($V_{i}$ in mev) or attenuation factor ($\alpha$ in$m^{-1}$). The
  Fermi energy $E_f=12.0 mev$. The length parameters are same as in
  FIG. 4.}
\end{figure*}

Having defined the total transmission as in Eq.~2, we now focus on the
resonant tunneling seen in such devices and the effect of absorption
on it. We first look into the effect of incoherence on the total
transmission (conductance in the linear response regime) in both
models of incoherence in the absence of finite bias. We see with
increase in incoherence in both models the width of transmission peak
increases (see FIG.~3) and height decreases. Also the transmission
in the valley region increases in qualitative agreement with prevalent
notions of decoherence. In this figure and in all subsequent figures
$\alpha=1e+05$ implies $1.0X10^{5}$.  We next focus on the total
transmission $G$, coherent transmission $T_{21}$, coherent reflection
$R_{11}$ and absorption defined as $A=1-T_{21}-R_{11}$, for both
Coherent and Wave Attenuation models of absorption. These are plotted
in FIG. 4.  We show that in contrast to Coherent Absorption wherein
absorption increases and then continuously decreases to
zero\cite{rubio}(in the extreme limit of $V_{i}\rightarrow\infty$ the
absorber acts as a perfect reflector), in case of Wave Attenuation $A$
increases and then saturates. One of the problems associated with a
coherent absorber is that absorption without reflection is not
possible and it introduces spurious scattering.  A note to be
added is that for wider barriers $A$ increases and then decreases (of
course saturating) with further increase in incoherence in case of
Wave Attenuation also, but this is due to the fact that the prompt
part in the reflection amplitude $r_1$ (see Fig.~1) which is
unaffected by dephasing is large, and this interferes with other waves
which leads to the non-monotonic behavior seen. This prompt part is
also present in case of Coherent Absorption, but unlike Wave
Attenuation the $A$ goes to zero in the asymptotic
regime($V_{i}\rightarrow\infty$,  while in case of Wave Attenuation
$A$ always saturates at a value however small. Apart from absorption
the true character of these models will be revealed in a comparison of
there conductance properties. In case of imaginary
potential the total conductance $(G)$ continuously decreases and
as $V_i\rightarrow\infty$, $G\rightarrow 0$ (this limit is not shown
in the graph as it goes beyond the scale used), while in-case of Wave
Attenuation as $\alpha\rightarrow\infty$, $G$ goes to a constant value
which depends on the Fermi-energy and other system parameters.  Thus
we see that the modeling of incoherence by Wave Attenuation is indeed
justified as we want the conductance to go to a classical value and
not that the conductance itself should vanish, as in the case of
Coherent Absorption model. This limiting case clearly indicates that
the Wave Attenuation model is more physical than the Coherent
Absorption model. In the above analysis it should be noted that, the
conductance formula used is valid in the linear response regime.
 
Another important quantity relevant to the study of electron transport
in these double barrier heterostructures is the tunneling current as a
function of bias. To determine the tunneling current we follow the
procedure adopted by Tsu and Esaki\cite{esaki,gaston}, who have
obtained in the zero temperature limit the following expressions for
tunneling current as a function of the applied bias- (i)for $V\ge
E_{f}$

\begin{eqnarray}
I_{t}=\frac{em^{*}}{2\pi^{2}\hbar^{3}}\int_{0}^{E_{f}} dE (E_{f}-E)T_{tot}
\end{eqnarray}

and (ii)for $V < E_{f}$-
\begin{eqnarray}
I_{t}=\frac{em^{*}}{2\pi^{2}\hbar^{3}}[V\int_{0}^{E_{f}-V} dE T_{tot}
+\int_{E_{f}-V}^{E_{f}} dE (E_{f}-E) T_{tot}]
\end{eqnarray}

In FIG.~5, we plot the tunneling current in terms of
(I=$I_{t}/{\frac{em^{*}}{2\pi^{2}\hbar^{3}}}$)versus applied bias for
both Coherent Absorption as well as Wave Attenuation models of
absorption. We see that Wave Attenuation models sequential
transport in a much better manner than Coherent Absorption. Whereas in
case of Coherent Absorption we see that for small increase in $V_i$
sequential contribution (the valley current) increases but for large
values of $V_i$ we see that the sequential contribution (or the valley
current) actually decreases. This is not so in-case of Wave
Attenuation, as sequential transport increases to a defined limit
and then saturates as expected.

With this affirmation of the suitability of Wave Attenuation to
express sequential transport in RTD's. We now use this model and
compare it with some experimental results, and check how far it is
accurate. In the paper of Sollner, et.al.\cite{sollner} wherein the
peak to valley ratio 6:1 was obtained, the experiment was performed at
25K with system barrier(AlGaAs) height and length of 230.0 mev and
$50\AA$, while well(GaAs) width is also $50\AA$. We can approximate
the relaxation time at 25K to be $10^{-13}$ secs and Fermi energy for
GaAs concentration of $1.0X10^{-18}m^{3}$ is $18.67 mev$. Thus
electron mean free path defined to be $V_{f}$ the Fermi velocity
multiplied by the relaxation time is $10^{-8}m$. $2\alpha$ defined as
inverse of mean free path is thus $0.5 X 10^{8} m^{-1}$. We find from
Eq.~4, the peak to valley ratio to be 3.5:1. Thus our model although,
not quite close to experimental results, nevertheless reflects
qualitatively the negative differential resistance in Resonant
tunneling diodes quite well. This peak-valley ratio is sensitive to
the relaxation time and in our analysis we have used the value
as quoted in Ref.\onlinecite{Yu}.

\section{Mesoscopic Rings}

\begin{figure} [h]
\protect\centerline{\epsfxsize=3.0in\epsfbox{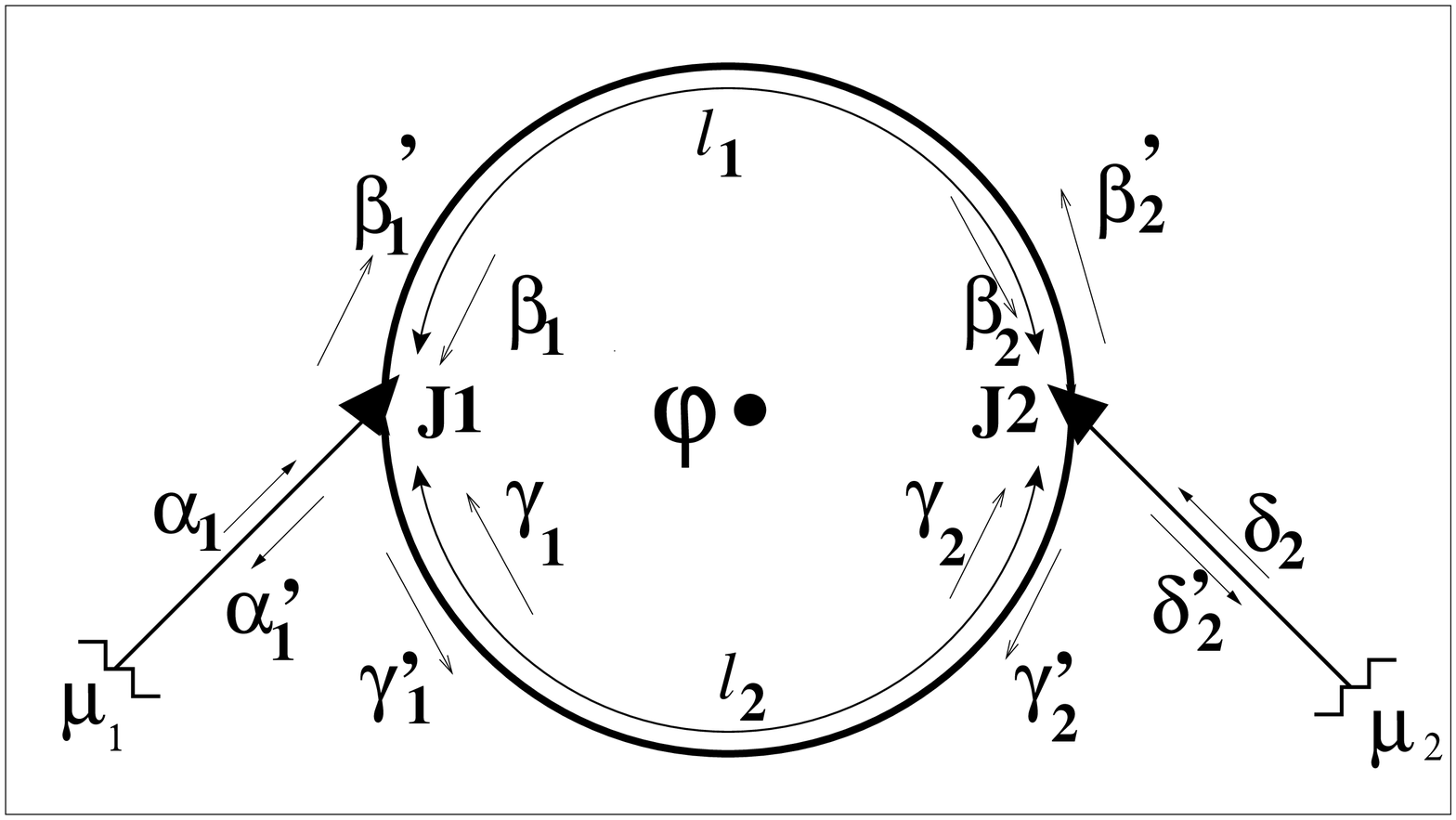}}
\caption{The asymmetric Aharonov-Bohm ring.}
\end{figure}

In this section we report on the extensively studied non-local
phenomena of Aharonov-Bohm oscillations\cite{webb,imry}. The system under
consideration is shown in FIG.~6. It is an asymmetric loop with upper
and lower arm lengths $l_1$ and $l_2$ and circumference
$L=l_{1}+l_{2}$, coupled to two leads which in turn are connected to
two reservoirs at chemical potentials $\mu_1$ and $\mu_2$. Inelastic
scattering is assumed to be absent in the leads while it is present in
the reservoirs, and in the loop we introduce incoherence via
Wave Attenuation or Coherent Absorption to simulate
inelastic scattering.  The S matrix for the left coupler yields the
amplitudes
$O_{1}=(\alpha_{1}^\prime,\beta_{1}^\prime,\gamma_{1}^\prime)$
emanating from the coupler in terms of the incident waves
$I_1=(\alpha_{1},\beta_{1},\gamma_{1})$, and for the right coupler
yields the amplitudes
$O_{2}=(\delta_{2}^\prime,\beta_{2}^\prime,\gamma_{2}^\prime)$
emanating from the coupler in terms of the incident waves
$I_2=(\delta_{2},\beta_{2},\gamma_{2})$. The S-matrix for either of
the couplers\cite{butipra} is given by-

\[S=\left(\begin{array}{ccc}
-(a+b)       & \sqrt\epsilon&\sqrt\epsilon\\
\sqrt\epsilon& a            &b            \\
\sqrt\epsilon& b            &a
\end{array} \right) \]

with $a=\frac{1}{2}(\sqrt{(1-2\epsilon)} -1)$ and
$b=\frac{1}{2}(\sqrt{(1-2\epsilon)} +1)$. Herein, $\epsilon$ plays the
role of a coupling parameter. The maximum coupling between reservoir
and loop is $\epsilon=\frac{1}{2}$, and for $\epsilon=0$, the coupler
completely disconnects the loop from the reservoir. We first consider
Wave Attenuation in which inelastic scattering in the arms of the
AB interferometer is taken into account by introducing an attenuation
constant per unit length in the two arms of the ring, i.e., the
factors $e^{-\alpha l_1}$ (or $e^{-\alpha l_2}$) in the  free
propagator amplitudes, every time the electron\cite{colin,datta}
traverses the upper (or lower) arms of the loop (see Fig.~1), as
discussed earlier.

The waves incident into the branches of the loop are related by the S
Matrices \cite{cahay}for upper branch by-

\[\left(\begin{array}{c}
\beta_1\\
\beta_2\\
\end{array} \right) \ =\left(\begin{array}{cc}
0     & e^{ikl_1} e^{-\alpha l_1} e^\frac{-i \theta l_1}{L}\\
e^{ikl_1} e^{-\alpha l_1} e^\frac{i \theta l_1}{L} & 0 \\
\end{array} \right) \left(\begin{array}{c}
\beta_1^\prime\\
\beta_2^\prime
\end{array} \right)\]
and  for lower branch-

\[\left(\begin{array}{c}
\gamma_1\\
\gamma_2\\
\end{array} \right) \ =\left(\begin{array}{cc}
0     & e^{ikl_2} e^{-\alpha l_2} e^\frac{i \theta l_2}{L}\\
e^{ikl_2} e^{-\alpha l_2} e^\frac{-i \theta l_2}{L} & 0 \\
\end{array} \right) \left(\begin{array}{c}
\gamma_1^\prime\\
\gamma_2^\prime
\end{array} \right)\]

These S matrices of course are not unitary $
S(\alpha)S(\alpha)^\dagger\neq 1$ but they obey the duality\cite{been}
relation $ S(\alpha)S(-\alpha)^\dagger= 1$. Here $kl_1$ and $kl_2$ are
the phase increments of the wave function in absence of flux.
$\frac{\theta l_1}{L}$ and $\frac{\theta l_2}{L}$ are the phase shifts
due to flux in the upper and lower branches.  Clearly, $\frac{\theta
  l_1}{L}+\frac{\theta l_2}{L}=\frac{2\pi\Phi}{\Phi_0} $, where $\Phi$
is the flux piercing the loop and $\Phi_0$ is the flux
quantum$\frac{hc}{e}$. The transmission and reflection coefficients
are given as follows-
$T_{21}=|\frac{\delta_{2}^\prime}{\alpha_{1}}|^2$,
$R_{11}=|\frac{\alpha_{1}^\prime}{\alpha_{1}}|^2$,
$R_{22}=|\frac{\delta_{2}^\prime}{\delta_{2}}|^2$,
$T_{12}=|\frac{\alpha_{1}^\prime}{\delta_{2}}|^2$ wherein wave
amplitudes $\delta_{2}^\prime,\delta_{2},\alpha_{1}^\prime,\alpha_{1}$
are as depicted in FIG.~6.

The coherent transmission coefficient $T_{21}$ from reservoir 1 to 2
is not symmetric under flux reversal and is due to the fact that
unitarity has been violated. For calculating the total conductance in
case of Coherent Absorption we use Eq.~2, and potential across the
ring is imaginary $V_{1}$ but we follow the wave guide
method\cite{xia,deo_stub}.  The waveguide
and S-Matrix methods are not exclusive, in fact the waveguide method
is just a special case of the S-Matrix, corresponding to a coupling
constant $\epsilon=4/9$.

In this method (see FIG.~6) the wave functions in the leads are-
$\Psi_{1}=e^{i k x}+r e^{-i k x}$ and $\Psi_{2}=t e^{i k x}$ for left
and right leads respectively. The wavefunctions in the upper and lower
arms of the loop are $\Psi_{u}=a_{u}e^{i k_{1}^{\prime} x}+b_{u} e^{-i
  k_{2}^{\prime} x}$ and $\Psi_{l}=a_{l} e^{i k_{2}^{\prime} x}+b_{l}
e^{-i k_{1}^{\prime} x}$. Here
$k_{1}^{\prime}=k^{\prime}-\frac{\phi}{L},
k_{2}^{\prime}=k^{\prime}+\frac{\phi}{L}$, with $k^{\prime}=k_{p}+i
k_{m}$ and
$k_{p}=\frac{k}{\sqrt{2}}\sqrt{1+\zeta},k_{m}=\frac{k}{\sqrt{2}}\sqrt{\zeta-1}
$, where $\zeta=\sqrt{1+(\frac{V_{i}}{k^2})^{2}}$.

\begin{figure*} 
\protect\centerline{\epsfxsize=6.0in\epsfbox{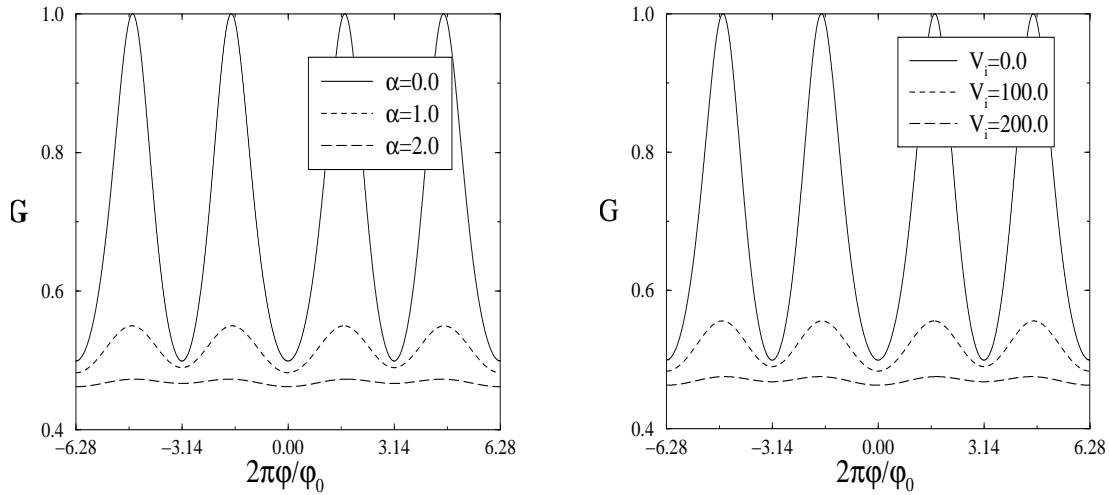}}
\caption{Comparison of two models of dephasing. Herein the
  Conductance (absorption in the inset) for the two cases Wave
  Attenuation(WA) and Coherent Absorption(CA) have been plotted. The
  parameters (all in dimensionless units) used are $kL=5.0,
  l_{1}/L=0.375, l_{2}/L=0.625, \Phi=1.0,$ and $ \epsilon=4/9$}
      \end{figure*}

\begin{figure} [h]
\protect\centerline{\epsfxsize=3.0in\epsfbox{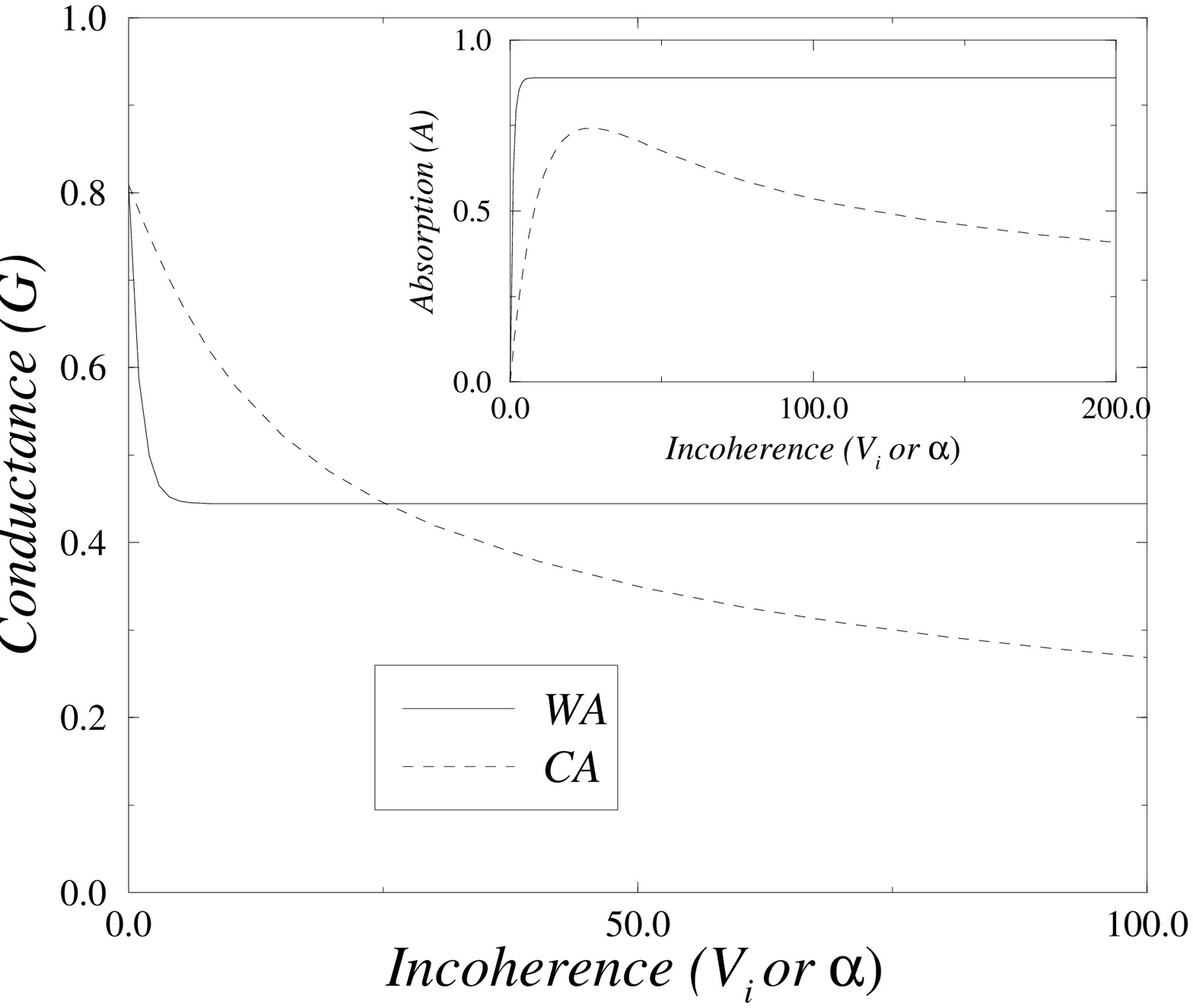}}
\caption{Comparison of two models of dephasing. Herein the
  Conductance (absorption in the inset) for the two cases Wave
  Attenuation(WA) and Coherent Absorption(CA) have been plotted. The
  parameters (all in dimensionless units) used are $kL=5.0,
  l_{1}/L=0.375, l_{2}/L=0.625, \Phi=1.0,$ and $ \epsilon=4/9$}
\end{figure}

Thus we have $R_{11}=|r|^{2}, T_{21}=|t|^{2}$. For electrons incident
from the right lead we adopt a similar procedure to calculate $R_{22}$
and $T_{12}$. Substituting these coefficients in Eq.~2, we get the
total conductance in case of Coherent Absorption. The total
conductance (G) in both the cases of absorption obeys Onsager's
symmetry while the coherent transmission contribution isn't. The
conductance shows flux periodicity of period $\Phi_0$ as expected for
ballistic rings\cite{buti,webb,gia}. In the ballistic and asymmetric
rings for certain values of the physical parameters $\Phi_0/2$
periodicity has also been seen experimentally and explained using the
single channel coherent transport model without dephasing. In our
present case as we scan the range of Fermi wavevector then in some
regions we observe $\Phi_0/2$ periodicities a much explored phenomena
in these systems as shown in Fig.~7. In both cases, i.e., Coherent
Absorption as well as Wave Attenuation, we see that increase in
incoherence leads to shifts of exact fermi wavevector wherein
$\Phi_0/2$ periodicities occur. Also the visibility defined as the
ratio of the difference between maximum and minimum conductances to
the sum of maximum and minimum conductances shows similar behavior in
either case. In both these cases visibility decreases as a function of
incoherence parameter signalling dephasing.  Moreover it has been
shown that\cite{condmat} $\Phi_0/2$ periodicity can also be observed
in the presence of incoherence which is of-course true for $\Phi_0$
periodicity. To truly bring out the differences between these models
of incoherence one has to study the total conductance and absorption
as a function of the incoherence parameters.

To this end, in FIG.~8 we plot the total conductance $(G)$ for both
the cases. In the case of Coherent Absorption we see that the total
conductance $(G)$ continuously decreases and as
$V_i\rightarrow\infty$, $G\rightarrow 0$ (this limit is not shown in
the graph as it goes beyond the scale used), while in-case of Wave
Attenuation as $\alpha\rightarrow\infty$, $G$ goes to a constant value
which depends on the Fermi-energy and other system parameters.  Thus
we see that the modeling of dephasing by Wave Attenuation is indeed
justified as we want the AB oscillations to die out and not that the
conductance itself should vanish, and this is where Wave Attenuation
scores over the Coherent Absorption model.  In the figure inset we
have depicted the behavior of total absorption in the system $A$ for
both the cases for the same physical parameters.  As
$\alpha\rightarrow\infty$ there is a finite absorption in the system
as the electron propagates in the medium in this limit whereas
absorption in the imaginary potential model is non-monotonic and in
the limit $V_i\rightarrow\infty$ absorption vanishes. This is due to
the fact that in the Coherent Absorption model as
$V_i\rightarrow\infty$ absorber acts as a perfect reflector as
mentioned before. There is no absorption in the medium as the
particles do not enter the medium (and hence $G=0$) obviously which is
an unrealistic situation for real systems. 

With our model for inelastic scattering we can also study
theoretically the effect of incoherence on Fano resonant line shapes.
This effect has earlier been seen in various contexts in mesoscopic
systems. A recent example being, Fano effect through a quantum dot in
an Aharonov-Bohm interferometer by Kobayashi, et.al\cite{kobay},
wherein they see this effect in the nature of the conductance as a
function of the gate voltage (Fermi energy).  Fano resonances (i.e.,
asymmetric line shapes) arise when a discrete set of states are
coupled to the continuum, these were first proposed by Ugo Fano to
explain the characteristic asymmetric line shapes in nuclear
scattering phenomena\cite{fano}. Basically it is the effect of
interference between two alternatives in which an electron can emerge
via resonant localized state or through the continuum. 
Earlier, in an analysis of a simple quantum dot system taking into
account the  contribution from  direct(non-resonant) path and
resonant path the form of the Fano resonance was shown to be 
 \cite{clerk}
\begin{eqnarray}
G(E)=G_{d}\frac{|2E+q\Gamma|^{2}}{4E^{2}+\Gamma^{2}}
\end{eqnarray}
In the above Eq.~5, $G=|t_{d}+t_r|^2$ is the conductance with
$t_d=e^{i\beta_{d}}\sqrt{G_d}$ and $t_{r}(E)=z_{r}\Gamma/(2E+i\Gamma)$
describing the direct and resonant transmission amplitudes, E the
energy, $\Gamma$ the resonance width, $G_{d}$ the non-resonant
conductance, and $q=i+z_{r}e^{-i\beta_{d}}/\sqrt{G_d}$ the complex
Fano parameter. It must be remembered that the above form of the Fano
line-shape is valid in the absence of dephasing.  In the presence of
dephasing the form of the total conductance $G$ is as given in Eq.~2.
It is claimed earlier \cite{clerk} that the second part of Eq.~2 will
lead to a Breit-Wigner line-shape in the presence of dephasing.  We
will study this above assertion in our present study of quantum dot in
an Aharonov-Bohm ring geometry which is different from that considered
earlier\cite{clerk} using Wave Attenuation method.

In the model we consider, a quantum dot
placed on the upper arm of an open Aharonov-Bohm ring. The quantum
dot has resonant discrete levels, the interplay of these states with
the continuous levels in the Aharonov-Bohm ring gives rise to Fano
resonances. We assume the discrete levels of the quantum dot to be
having Breit-Wigner forms.  Following the formalism of Sun and Lin
\cite{lin}, one can write the $pq^{th}$ element of the 2 X 2 S-Matrix
of the quantum dot as-

\begin{eqnarray}
S_{pq}=\delta_{pq}-\frac{i/2}{\big[\sum_{j}\frac{\Gamma_{j}}{E-E_{j}}\big]^{-1}+i/2} 
\end{eqnarray}

wherein $\Gamma_{j}$ and $E_j$ represent the width and energy of the
$j^{th}$ resonant level of the quantum dot while $E$ denotes the Fermi
energy and $p,q$ take on values $1$ and $2$. It should, however, be
noted that unlike Sun and Lin, we neglect the intra-dot Coulomb
interaction as in Eq.~6 and solve the scattering problem by placing
the quantum dot symmetrically in the upper arm of the original
Aharonov-Bohm ring(FIG.~6).

\begin{figure} [t]
\protect\centerline{\epsfxsize=3.0in\epsfbox{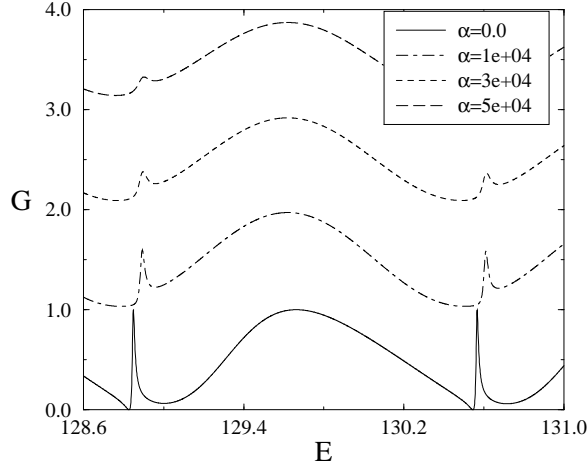}}
\caption{The Conductance (G) $vs.$ E in $mev$. Assuming dot has 5 states with
  $\Delta E=20$, $E_{j}=(2j-1)10,$ $\Gamma_{j}$ dependent on state $j$
  with $\Gamma_{1}=5.0$ and $\Gamma_{j}=1.1\Gamma_{j-1}$. The
  Conductance for increasing values of $\alpha$ in $m^{-1}$ are
  shifted by 1.0 for clarity.  The length parameters of the
  Aharonov-Bohm ring are $l_{1}=l_{2}=10\AA$ and the dot is
  symmetrically placed on the upper arm. The coupling strength is 0.5
  (maximal) and flux is $0.3$ $weber$.  The ring as in the experiment
  is assumed to be of $GaAs$.}
\end{figure}

In FIG.~9 we plot the Conductance as a function of incident electron
energy(in a narrow range) using Eq.~2 for various values of incoherent 
parameter $\alpha$. The Conductance shows Fano line-shapes as one
varies the incident electron energy. In this
figure the physical parameters are as mentioned in the figure caption.  As
inelastic scattering parameter $\alpha$ is increased Fano line-shapes broaden
and their peak values are reduced. Subtracting the continuum
background of the Conductance, we see that the asymmetric Fano
line-shapes evolve to symmetric line-shapes corresponding to Lorentzian
or Breit-Wigner forms in similarity with experimental observations and 
also consistent with the treatment of Ref.\onlinecite{clerk}.
The incoherence parameter $2\alpha$ is defined as equal to inverse of
the mean free path as in the previous section. A representative value
of $\alpha$ in FIG.~9, for example $1X10^{4}$ corresponds to a mean
free path of $5 X 10^{-5} m $ which is in conformity with the
millikelvin temperature range at which the experiment has been
performed. The complete understanding requires details of physical
parameters associated with the real experiment.  The complex Fano
parameter as defined above also changes with increase in dephasing
strength, as seen in Ref.\onlinecite{clerk}. Another feature as seen
in the experiment is the transition as a function of applied flux, of
an asymmetric line-shape to symmetric and then back to asymmetric,
further work on these is currently on and will be reported elsewhere.

\section{Conclusions}

In this work we have brought out some of the unphysical
characteristics associated with the Coherent Absorption model as
compared to the Wave Attenuation model which have been used earlier to
study dephasing in mesoscopic systems. Both these models are
phenomenological in nature and are applied to study transport across
Resonant tunneling diodes and Aharonov-Bohm rings. Further we have
studied the peak to valley ratios in case of Resonant tunneling
diodes, dephasing of magneto-conductance oscillations and Fano
resonances. We have shown that the Wave-Attenuation model is
consistent with prevalent notions of decoherence.


\begin{thebibliography} {99}

\bibitem{gas} S. Gasiorowicz, {\it Quantum Physics}
 (Wiley Eastern, New York, 2000).

\bibitem{schiff} L. I. Schiff, {\it Quantum Mechanics}
 (Mc Graw Hill, New York, 1974).

\bibitem {ferry} D. Ferry and J. Barker, \apl {\bf 74}, 582 (1999).

\bibitem {stone} A. D. Stone and P. A. Lee, \prl {\bf 54}, 1196 (1985).

\bibitem {pradhan} P. Pradhan and N. Kumar, \prb {\bf 50}, R9644
  (1994).

\bibitem{joshi} Sandeep K. Joshi, D. Sahoo and A. M. Jayannavar, \prb
    {\bf 62}, 880 (2000).  

\bibitem {colin} Colin Benjamin and A. M. Jayannavar, \prb {\bf 65},
  155309 (2002).

\bibitem{soj1} S. Anantha Ramakrishna and N. Kumar, preprint
  cond-mat/0009269.

\bibitem{soj2} C. Benjamin and A. M. Jayannavar, \ssc {\bf 121}, 591 (2002).

\bibitem {neutron} H. Rauch and J. Summhammer, \pra {\bf 46}, 7284 (1992).

\bibitem {land} R. Landauer, \pm {\bf 21}, 863 (1970).

\bibitem {imry}  Y. Imry, {\it An introduction to mesoscopic physics}
 (Oxford, 2001).

\bibitem {buti} M. B\"{u}ttiker, \prb {\bf 33}, 3020 (1986);\ibmjrd   {\bf
    32}, 63 (1988).
  
\bibitem {buti1} M. B\"{u}ttiker in {\it Resonant Tunneling in
    Semiconductors, Physics, and Applications}, edited by L. L. Chang,
  E. E. Mendez and C. Tejedor (Plenum , New York, 1991), p.213.

  
\bibitem {jayan} A. M. Jayannavar, \prb {\bf 49}, 14718 (1994); A. K.
  Gupta and A. M. Jayannavar, \prb {\bf 52}, 4156 (1995); S. K.Joshi
  and A. M. Jayannavar, \ijmp {\bf 12}, 1555 (1998); S. K.Joshi and
  A. M. Jayannavar, \ijmp {\bf 14}, 1669 (2000).



\bibitem {mortensen} N. A. Mortensen, A. P. Jauho and K.
     Flensberg, Superlattices and Microstructures {\bf 28}, 67 (2000).    

\bibitem {datta} S. Datta, {\it Electron Transport in mesoscopic systems}
   (Cambridge University press, Cambridge, 1995).

\bibitem {brouwer} P. W. Brouwer and C. W. J. Beenakker, \prb {\bf
      55}, 4695 (1997); P. W. Brouwer, Ph.D. thesis, Insttuut-Lorentz,
    University of Leiden, The Netherlands,1997.


\bibitem{gaston} G. Garcia-Calderon, in {\it The Physics of
    Low-Dimensional Semiconductor Structures}, edited by Butcher
  et. al. (Plenum, New York, 1993), p.267.

\bibitem{rubio} A. Rubio and N. Kumar, \prb {\bf 47}, 2420 (1993).

\bibitem {esaki}R. Tsu and L. Esaki, \apl {\bf 22}, 562 (1973).

\bibitem {chang} L. L. Chang, L. Esaki and R. Tsu, \apl {\bf 24}, 593 (1974).

\bibitem {sollner} T. C. L. G. Sollner, W. D. Goodhue, P. E. Tannewald, 
  C. D. Parker and D. D. Peck, \apl {\bf  43}, 588 (1983).

\bibitem {Yu} Hu Yuming, \jpc {\bf 21}, L23 (1988).

\bibitem {griffith} D. J. Griffiths, \ajp {\bf 72}, 3584 (2001).

\bibitem {azbel}B. Ricco and M. Ya. Azbel, \prb {\bf 29}, 1970 (1984).


\bibitem {grincwajg} M. Johnson and A. Grincwajg , \apl {\bf 51}, 1729
  (1987).
  

\bibitem {zohta} Y. Zohta and H. Ezawa, \jap {\bf 72}, 3584 (1992).

\bibitem{pareek} T. P. Pareek, S. K. Joshi and A. M. Jayannavar, \prb
    {\bf 57}, 8809 (1998).

\bibitem{lan_romp} R. Landauer and Y. Imry, \rmp {\bf 71}, S306 (1999).

\bibitem {butiparama}  M. B\"{u}ttiker, \pjp {\bf 58}, 241 (2002).

\bibitem{webb} S. Washburn and R. Webb, Adv. Phys.  {\bf 35} 375  (1986).

\bibitem {butipra} M. B\"{u}ttiker, Y. Imry and M. Ya. Azbel, \pra {\bf
    30}, 1982 (1984).
     
\bibitem {cahay} M. Cahay, S. Bandopadhyay and H. Grubin, \prb {\bf
       39}, R12989 (1986).

\bibitem {been} J. C. J. Paasschens, T. Sh. Mishirpashaev and
  C. W. J. Beenakker, \prb{\bf 54}, 11887 (1996).
  
\bibitem{xia}J. B. Xia, \prb {\bf 45}, 3593 (1992).
  
\bibitem{deo_stub} P. S. Deo and A. M. Jayananvar, \prb {\bf 50},
  11629 (1994); A. M. Jayananvar and P. S. Deo, \prb {\bf 49}, 13685
  (1994); T. P. Pareek, P. S. Deo and A. M. Jayananvar, \prb {\bf
    52}, 14657 (1995); Sandeep K. Joshi, D. Sahoo and A. M.
  Jayannavar, \prb {\bf 64}, 075320 (2001).


\bibitem{gia} Y. Gefen, Y. Imry, M. Ya. Azbel, \prl {\bf 52}, 129 (1984).


\bibitem{condmat} C. Benjamin, S. Bandyopadhyay and A. M. Jayannavar,
  preprint cond-mat/0205662, \ssc (2002) in press.

\bibitem{kobay} K. Kobayashi, H. Aikawa, S. Katsumoto and Y. Iye, \prl
  {\bf 88}, 256806 (2002). 

\bibitem{fano} U. Fano, Phys. Rev. {\bf 124}, 1866 (1961).

\bibitem{clerk} A. A. Clerk, X. Waintal and P. W. Brouwer, \prl
  {\bf 86}, 4636 (2001). 

\bibitem{lin}Q. Sun and T. Lin, \epjb {\bf 5}, 913 (1998).

\end{thebibliography}
\end{document}